# A Method to Calculate Correlation for Density Functional Theory on a Quantum Processor

Ryan Hatcher, Jorge A. Kittl and Christopher Bowen

*Advanced Logic Laboratory, Samsung Semiconductor Inc., Austin, Texas, 78754 USA*

An extension of the Variational Quantum Eigensolver (VQE) method is presented where a quantum computer generates an accurate exchange-correlation potential for a Density Functional Theory (DFT) simulation on classical hardware. The method enables efficient simulations of quantum systems by interweaving calculations on classical and quantum resources. DFT is implemented on classical hardware, which enables the efficient representation of and operation on quantum systems while being formally exact. The portion of the simulation operating on quantum hardware produces an accurate exchange-correlation potential but only requires relatively short depth quantum circuits.

## I. INTRODUCTION

Since the quantum computer was first proposed, numerous quantum algorithms have been developed to solve many different kinds of problems [1]. However, it may be that the first practical application of quantum computing could be to solve that first problem for which it was proposed: the simulation of many-body quantum systems [2]. Subsequent approaches have been proposed to efficiently simulate the exponential state space of both time-dependent and time-independent quantum systems [3, 4]. However, quantum computing platforms available to date have imposed significant constraints on simulations due to both the limited number of qubits available for the calculations as well as the relatively short qubit coherence times and modest gate fidelity. Therefore, a useful quantum computing application must be able to demonstrate superiority (i.e. quantum supremacy) over a classical simulation using relatively few qubits with a short depth. For example, Quantum Phase Estimation (QPE) [5, 6, 7] can be employed to find the ground state or even excited states of quantum systems but requires a machine that can maintain coherence for many more sequential gate operations than have been achievable in practice for nontrivial problems. The goal for the proposed approach is to harness the ability of quantum computers to efficiently represent and operate on the exponential state space of a many-body quantum state without relying on deep quantum circuits for which no hardware is available nor likely to be available for many years or even decades.

To that end, a hybrid quantum/classical simulation of quantum mechanics is proposed where the strengths of each computational platform offsets the weakness of its counterpart. This approach is an extension of the Variational Quantum Eigensolver (VQE) method [8, 9], which has been demonstrated for small molecules. Unlike previous VQE instantiations, the classical portion of the simulation proposed here is largely executed within a Density Functional Theory (DFT) framework along with some limited pre- and post-processing. DFT is an attractive vehicle for the classical portion of the simulation given the tremendous existing infrastructure that can be leveraged. In addition, the DFT formalism is both exact and tractable to simulate quantum mechanics on classical hardware [10]. The DFT formalism casts the usual many-body problem where the state space is exponential in the number of particles being simulated instead to a system of non-interacting particles that can be described by a set of single-particle states that grows linearly with the number of particles being simulated. All the many-body effects are swept into a single approximation, known as the exchange-correlation energy, which is a functional of the density alone. Although the exact exchange-correlation functional is not known, approximations have improved significantly over the last 70+ years and proven to be surprisingly accurate for many systems (see [11] for an overview of the depth and breadth of exchange-correlation functionals). Still, DFT fails spectacularly for a number of systems of interest [12] and while improvements to exchange up to and including exact exchange can be introduced, in general it is not tractable to systematically correct errors in the approximation to correlation effects. In the proposed approach, the quantum portion of the simulation provides the classical portion with just such a systematic means to include accurate correlation effects via a corrected exchange-correlation functional.

## II. BACKGROUND

In this section, relevant quantities from the second quantization formalism are defined and a brief overview of the VQE method is presented. The second quantized form of the time-independent Schrodinger equation is defined in a basis of orthonormal single particle functions, $|\psi_i\rangle$. An $N$-particle state corresponding to the single particle basis with $M$ basis functions can be described by the occupation number vector:

$$|\vec{n}\rangle = |n_1, n_2, \ldots, n_M\rangle \quad (1)$$

Each labelled occupation, $n_i$, corresponds to a single particle basis function, $\psi_i$, and is either unoccupied or occupied by one particle, $n_i = \{0, 1\}$. For an $N$-particle state there is the constraint:

$$\sum_{i=1}^{M} n_i = N \qquad (2)$$

The many-body wavefunction can be constructed from a linear combination of $P$ states in the occupation basis.

$$|\Psi\rangle = \sum_{\alpha=1}^{P} c_\alpha |\vec{n}^\alpha\rangle \qquad (3)$$

For $N$ particles in a basis of $M$ states, the number of terms in the many-body wavefunction is:

$$P = \frac{M!}{N!(M-N)!} \qquad (4)$$

Define the one-body reduced density matrix:

$$\rho_{ij} \equiv \langle \Psi | \hat{a}_i^\dagger \hat{a}_j | \Psi \rangle \qquad (5)$$

$\hat{a}_i^\dagger$ ($\hat{a}_j$) are creation (annihilation) operators. The two-body reduced density matrix is defined as:

$$\Gamma_{ijkl} \equiv \langle \Psi | \hat{a}_i^\dagger \hat{a}_k^\dagger \hat{a}_l \hat{a}_j | \Psi \rangle \qquad (6)$$

The second quantized Hamiltonian for a set of $N$ electrons in an external field $\hat{V}_{ext}$ is:

$$\hat{H} = \sum_{i,j=1}^{M} (t_{ij} + v_{ij}^{ext}) \hat{a}_i^\dagger \hat{a}_j +$$
$$\frac{1}{2} \sum_{i,j,k,l=1}^{M} v_{ijkl}^{ee} \hat{a}_i^\dagger \hat{a}_k^\dagger \hat{a}_l \hat{a}_j \qquad (7)$$

$t_{ij}$ is the kinetic energy matrix element, $v_{ij}^{ext}$ is the single-particle external potential matrix element (e.g. due to nuclei or an external magnetic field) and $v_{ijkl}^{ee}$ is the electron-electron potential matrix element (see Appendix B).

Given a wavefunction, the energy is the expectation value of the Hamiltonian, which can be expressed as the trace of the reduced density matrices:

$$\mathcal{E} = \sum_{i,j=1}^{M} (t_{ij} + v_{ij}^{ext}) \rho_{ij} + \frac{1}{2} \sum_{i,j,k,l=1}^{M} v_{ijkl}^{ee} \Gamma_{ijkl} \qquad (8)$$

Aspuru-Guzik et al [8] first proposed the VQE method where the single-particle matrix elements are calculated classically and the reduced one- and two-body density matrices are calculated on a quantum computer. The wavefunction is not known explicitly but instead prepared on the quantum computer by applying a sequence of parametrized operations on a well-defined reference state. Note that the parameters are essentially the settings on the "knobs" of the machine that can be stored classically. To find the ground state, a classical optimization algorithm is applied to the parameters where the objective function is to minimize the total energy. The expectation value of the Hamiltonian can be calculated for a trial wavefunction by mapping the fermionic creation/annihilation operators to operators that can be implemented directly in hardware. Fortunately, gate-based quantum computers support Pauli spin operators directly and several methods to efficiently map fermionic annihilation/creation operators to Pauli operators have been proposed and demonstrated [9, 13, 14, 15, 16]. Each of the one- and two-body reduced density matrix elements (5) and (6) can then be cast as a sum of tensored single-qubit Pauli operators and can be estimated by measuring the expectation value of the individual terms qubit by qubit.

The process is repeated iteratively with a new guess for the set of wavefunction parameters from a classical minimization algorithm. This approach has the advantage of enabling calculations of the total energy that formally include the exchange and correlation effects of a second quantized many-body wavefunction. Moreover, only relatively short-depth quantum circuits are required to calculate each term in the expectation value. Finally, although one doesn't have access to the wavefunction directly, the parameters to prepare the lowest energy wavefunction are available and thus any property can be calculated for which there exists a known operator.

## III. METHOD

A hybrid approach is described below that is an extension of the VQE method. The key difference is that the method maintains both a set of non-interacting wavefunctions on a classical platform as well as a set of interacting wavefunctions on a quantum platform that both map to the *same* density upon convergence. This enables one to take full advantage of the DFT formalism on the classical side, where it is tractable to represent and operate on a set of non-interacting wavefunctions. It is an iterative method where calculations on a classical platform are interweaved with calculations on quantum hardware.

DFT provides an exact treatment of the many-body quantum ground state given the exact exchange-correlation energy functional, $E_{xc}[\rho]$ [10]. The advantage of DFT on classical platforms is that it requires only the storage of and operation on the electronic density and a small set of fictitious non-interacting wavefunctions, which can be stored efficiently. Once a ground state DFT simulation has converged, any ground state quantity for which an operator can be defined can be calculated relatively efficiently as long as the exact $E_{xc}$ is provided and the DFT density is equal to the many-body ground state density. However, only approximations to $E_{xc}$ have been available and DFT simulations tend to perform poorly for systems where correlation effects are significant [12].

There is an exponential increase in computation enabled by a quantum computer in the size of the many-body basis in which the Hamiltonian is diagonalized. Consider a many-body system with $N$ electrons occupying a subset of $M$ total states ($N \leq M$) where the number of possible occupation number vectors for such a system is given by (4). In the second quantized formulation, one can represent the occupation of each state with a qubit thereby efficiently encoding all $2^M$ possible occupations for $N = 0..M$ particles with only $M$ qubits. Even for quantum computers with a relatively modest number of qubits, there can be a substantial computational advantage. For example, a quantum computer with 50 qubits can represent the state space of a system with 25 fermions in 50 states efficiently, which corresponds to $1.26 \times 10^{14}$ terms in the many-body wavefunction. Furthermore, as described above, provided the matrix elements of the Hamiltonian from a classical simulation it is possible to efficiently map the second quantized Hamiltonian to Pauli operators that are then applied to a many-body wavefunction with an exponential number of occupation states. Following the VQE method, a classical optimization algorithm is then applied to find the parameters that generate a wavefunction that minimizes the total energy.

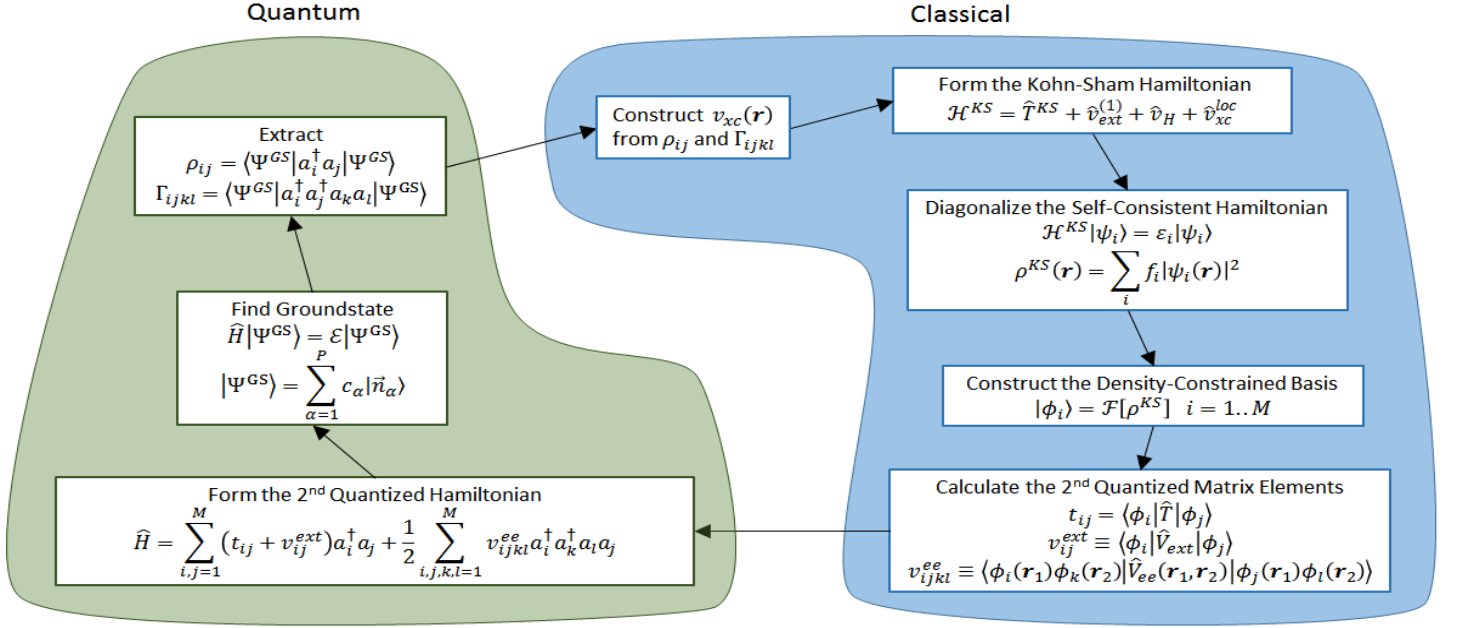

Figure 1. Flow diagram of the proposed scheme. It is an iterative scheme with the classical portion of the calculation shown on the right and the quantum tasks on the left. The corrected KS kinetic energy operator, $\hat{T}$, and the matrix elements in the corrected local exchange-correlation potential $\hat{v}_{xc}^{loc}$ are given in Appendix B.

### A. Tasks for the Classical Platform

There are three steps in the classical portion of the loop. The first step is to construct the exchange-correlation potential. The simulation is initiated by choosing an approximation to the exchange-correlation potential [11]. For all subsequent iterations, there is a set of reduced one- and two-body matrix elements that are passed in from the quantum portion of the simulation. The exchange-correlation potential is then constructed directly. Appendix B contains the derivation of and expression for the many-body corrected exchange-correlation potential (B48).

The second step is a DFT simulation. First, a non-interacting Hamiltonian is constructed as the sum of the kinetic energy operator and an effective potential.

$$\hat{\mathcal{H}}^{KS} = \hat{T}^{KS} + \hat{v}_{eff}^{KS} \qquad (9)$$

The Kohn-Sham (KS) Hamiltonian is then diagonalized in order to find the eigenvectors and eigenvalues, which are the KS wavefunctions, $\psi_i^{KS}$ and energies $\varepsilon_i^{KS}$.

$$\hat{\mathcal{H}}^{KS} \psi_i^{KS} = \varepsilon_i^{KS} \psi_i^{KS} \qquad (10)$$

Based on the KS eigenvalues, $\varepsilon_i^{KS}$, the occupations, $f_i$, of the KS wavefunctions are then calculated with which the density, $\rho^{KS}(r)$, can be constructed.

$$\rho^{KS}(r) = \sum_i f_i |\psi_i^{KS}(r)|^2 \qquad (11)$$

Although the KS wavefunctions and eigenvalues are fictitious, the density is exact in the limit that the exchange-correlation functional, $E_{xc}$, is exact ($\hat{v}_{eff}^{KS}$ is constructed in part from the $E_{xc}$).

The third step is to form a single particle basis set, $\phi_i$, with which the interacting wavefunctions will be constructed in the quantum portion of the loop and calculate the Hamiltonian matrix elements associated with this new basis. One option would be to use the KS states as the basis. However, the many-body density could be quite different from the DFT density for such a basis and this may prevent convergence between the quantum and classical portions of the calculation. Instead, the single particle basis is constructed so that each state is constrained to reproduce the density calculated from the occupied KS orbitals:

$$|\phi_i(r)|^2 = \frac{1}{N}\rho^{KS}(r) \qquad (12)$$

Since the volume integral of the density in (11) is equal to the number of electrons, $N$, the factor of $\frac{1}{N}$ in (12) ensures that $\langle \phi_i | \phi_i \rangle = 1$. Consider the following form of such a basis:

$$\phi_i(r) = \sqrt{\frac{\rho^{KS}(r)}{N}} e^{i\xi_i(r)} \qquad (13)$$

$\xi_i(r)$ is a real function that enforces orthonormality between the states [17, 18, 19]. Several bases of this form have been proposed [19, 20, 21]. The basis proposed by Zumbach and Maschke [20] is presented in Appendix A along with functional derivatives required to construct the exchange-correlation potential. The exchange-correlation potential is derived in Appendix B.

The last part of the classical portion of the calculation is to form and store the matrix elements that are needed to form the many-body Hamiltonian corresponding to this basis. Note that the density-constrained basis will not necessarily span the space of the occupied KS states.

### B. Tasks for the Quantum Platform

The goal of the quantum portion of the simulation is to take as input the single particle basis set, $\phi_i$, and corresponding matrix elements, $t_{ij}$, $v_{ij}^{ext}$, and $v_{ijkl}^{ee}$, from the classical calculation, find the lowest energy eigenstate. The one- and two-body reduced density matrices, $\rho_{ij}$ and $\Gamma_{ijkl}$, are then passed back to the classical portion of the simulation. Given a single-particle basis, $\phi_i$, that is both orthonormal and whose square modulus is the Kohn-Sham density, the density of any $N$ particle many body wavefunction has the following form:

$$\rho(r) = \big(1 + \Delta\rho(r)\big)\rho^{KS}(r) \qquad (14)$$

$\Delta\rho$ is the exact expression for the many-body correction to the KS density, which depends on the exact form of the single-particle basis set. For the constrained basis of the form given in (13), the many-body correction (from appendix B) is:

$$\Delta\rho(r) \equiv \frac{2}{N}\sum_{i<j}^{M} Re\left(\rho_{ij} e^{i(\xi_j - \xi_i)}\right) \qquad (15)$$

Note that by constraining the basis to satisfy (12), the many-body correction to the density will have an oscillatory term that should result in significant cancellation in the limit of large $M$. Furthermore, in the limit that the exact $v_{xc}$ is applied, the correction term will vanish identically since the KS density and the many-body density will be equal. It is conjectured though not proven here that as the simulation progresses, the many-body correction to the density (15) be reduced and the approximation to the exchange-correlation potential will improve as the big loop converges. It may even be that the magnitude of the density correction could be a reasonable quantity to test for convergence. Or perhaps the density correction itself can be cleverly leveraged to improve convergence directly.

## IV. DISCUSSION

An extension of the VQE method has been described where a quantum computer efficiently generates an accurate exchange-correlation potential for a DFT simulation on classical hardware. Both the VQE method and the method described here seek to enable accurate and efficient simulations of quantum systems in a complementary fashion by leveraging the strengths and offsetting the weaknesses of quantum and classical computing platforms. The key difference is that the method described here invokes the DFT formalism in the classical portion of the simulation, which enables the efficient representation of and operation on quantum systems while being formally exact. Moreover, this method harnesses significant functionality and infrastructure of existing DFT codes. The weakness of DFT is that many-body correlation effects must be approximated and there are no known tractable methods to improve these approximations systematically. On the quantum side, the weakness of the DFT simulation is mitigated by implementing a second quantized formalism enabling the efficient representation of and operation on an exponential number of terms in the occupation basis, which in turn enables the efficient calculation of many-body correlation effects. The weakness on the quantum side is that the hardware currently available (and perhaps for the foreseeable future) is limited to a modest number of qubits and the coherence times and gate fidelities limit the depth of the quantum circuits that can be executed. Interweaving classical calculations reduces the depth of the circuits that must be performed during the quantum portion thus mitigating the limitation on the depth that the quantum computer needs to support.

It may seem unnecessary and perhaps even counterproductive to construct the constrained single-particle basis from the DFT density. As noted in the previous section, it is conjectured that as the exchange-correlation potential improves, the many-body correction (15) to the density will get smaller. The justification for constructing a constrained basis is that it should always result in a relatively small many-body correction due to the oscillatory term in the sum in and should help speed up convergence in the overall calculation. However, there is no guarantee that the constructed basis will span the space of the original DFT eigenstates, nor is it guaranteed that this basis will span the space required for the system. One test for the quality of the constructed basis is to compare the ground state energy calculated in this single-particle basis against the energy of the ground state calculated with a single-particle basis of DFT eigenstates.

*Acknowledgements*.

The authors would like to thank Dr. Jay Gambetta and Dr. Kristen Temme for fruitful discussions.

# Appendix A    Zumbach Maschke Basis and Functional Derivative

Zumbach and Mashke (ZM) [20] proposed a density constrained basis with the form given by (13):

$$\phi_i(r) = \sqrt{\frac{\rho(r)}{N}} e^{i\xi_i(r)} \tag{A1}$$

The phase factor, $\xi_i$, for the $i^{th}$ basis vector is real-valued and enforces orthogonality between different basis states:

$$\xi_i(r) = k_i \cdot f(r) \tag{A2}$$

$k_i$ is a 3-vector of integers, $k_i^{x,y,z} \in \{0, \pm 1, \pm 2, \ldots\}$, that uniquely specify a basis state. $f(r)$ is a 3-vector function over space that is the same for all the basis states:

$$f(r) = \frac{2\pi}{N} \int_{x_1}^{x} dx'\, \bar{\bar{\rho}}(x')\, \hat{x} + \frac{2\pi}{\bar{\bar{\rho}}(x)} \int_{y_1}^{y} dy'\, \bar{\rho}(x, y')\, \hat{y} + \frac{2\pi}{\bar{\rho}(x,y)} \int_{z_1}^{z} dz'\, \rho(x, y, z')\, \hat{z} \tag{A3}$$

The limits of integration are $(x_1, x_2), (y_1, y_2), and\ (z_1, z_2)$. The planar and linear densities are defined as:

$$\bar{\bar{\rho}}(x) \equiv \iint_{y_1, z_1}^{y_2, z_2} \rho(x, y, z) dy\, dz \tag{A4}$$

$$\bar{\rho}(x, y) \equiv \int_{z_1}^{z_2} \rho(x, y, z)\, dz \tag{A5}$$

Note there is freedom to choose the orientation of the axes and for every orientation there are six independent permutations.

This basis is normalized by construction. To show that it is orthogonal consider the overlap for two states $k_i \neq k_j$.

$$\langle \phi_i | \phi_j \rangle = \frac{1}{N} \int dr\, \rho(r) e^{i(k_j - k_i) \cdot f(r)} \tag{A6}$$

Defining $k_{ji} \equiv k_j - k_i$ and substituting in (A3)

$$\langle \phi_i | \phi_j \rangle = \frac{1}{N} \int dx\, e^{ik_{ji}^x f_x(x)} \int dy\, e^{ik_{ji}^y f_y(x,y)} \int dz\, \rho(x, y, z) e^{ik_{ji}^z f_z(x,y,z)} \tag{A7}$$

It can be shown that the integral over z is:

$$\int dz\, \rho(x, y, z) e^{ik_{ji}^z f_z(x,y,z)} = \begin{cases} \bar{\rho}(x, y) &, k_{ji}^z = 0 \\ 0 &, k_{ji}^z \neq 0 \end{cases} \tag{A8}$$

The integral over y is:

$$\int dy\, \bar{\rho}(x, y)\, e^{ik_{ji}^y f_y(x,y)} = \begin{cases} \bar{\bar{\rho}}(x), & k_{ji}^z = k_{ji}^y = 0 \\ 0, & \text{else} \end{cases} \tag{A9}$$

Finally the integral over x is:

$$\frac{1}{N} \int dx\, \bar{\bar{\rho}}(x) e^{ik_{ji}^x f_x(x)} = \begin{cases} 1 &, k_{ji} = 0 \\ 0 &, k_{ji} \neq 0 \end{cases} \tag{A10}$$

The functional derivative of the phase factors will be required to derive the exchange correlation matrix elements in Appendix B. Note that the following derivation works for the phase factor of a single state $\xi_i$ or the product form, $\xi_{ji}$. The functional derivative of the general form of an exponentiated function is:

$$\frac{\delta e^{i\xi_{ji}(r')}}{\delta \rho(r)} = e^{i\xi_{ji}(r')} \frac{\delta \xi_{ji}(r')}{\delta \rho(r)} \tag{A11}$$

The functional derivative of the ZM phase factor is

$$\frac{\delta \xi_{ji}(r')}{\delta \rho(r)} = \frac{2\pi k_{ji}^x}{N} \frac{\delta}{\delta \rho(r)} \int_{x_1}^{x'} dx''\, \bar{\bar{\rho}}(x'') + \frac{\delta}{\delta \rho(r)} \frac{2\pi k_{ji}^y}{\bar{\bar{\rho}}(x')} \int_{y_1}^{y'} dy''\, \bar{\rho}(x', y'') + \frac{\delta}{\delta \rho(r)} \frac{2\pi k_{ji}^z}{\bar{\rho}(x',y')} \int_{z_1}^{z'} dz''\, \rho(x', y', z'') \tag{A12}$$

The first term is trivial to evaluate observing that:

$$\int_{x_1}^{x'} dx'' \frac{\delta \bar{\rho}(x'')}{\delta \rho(r)} = \int_{x_1}^{x'} \int_{y_1}^{y_2} \int_{z_1}^{z_2} dr'' \frac{\delta \rho(r'')}{\delta \rho(r)} = \int_{x_1}^{x'} \int_{y_1}^{y_2} \int_{z_1}^{z_2} dr'' \delta(r - r'') = \Theta(x' - x) \quad (A13)$$

The Heaviside function is defined as:

$$\Theta(x) = \begin{cases} 1, & x \geq 0 \\ 0, & x < 0 \end{cases} \quad (A14)$$

The second term requires the quotient rule for functional derivatives:

$$\frac{\delta}{\delta \rho(r)} \frac{\int_{y_1}^{y'} dy'' \bar{\rho}(x',y'')}{\bar{\rho}(x')} = \frac{1}{\bar{\rho}(x')} \frac{\delta}{\delta \rho(r)} \int_{y_1}^{y'} dy'' \bar{\rho}(x', y'') - \frac{\int_{y_1}^{y'} dy'' \bar{\rho}(x',y'')}{\bar{\rho}^2(x')} \frac{\delta \bar{\rho}(x')}{\delta \rho(r)} \quad (A15)$$

Given that any function, $g(x)$, can be written as a functional $g(x) = \int dx' g(x') \delta(x - x')$, the functional derivatives above can be evaluated as:

$$\frac{\delta}{\delta \rho(r)} \int_{y_1}^{y'} dy'' \bar{\rho}(x', y'') = \int_{x_1}^{x_2} \int_{y_1}^{y'} \int_{z_1}^{z_2} dr'' \frac{\delta \rho(r'')}{\delta \rho(r)} \delta(x' - x'') = \delta(x' - x) \Theta(y' - y) \quad (A16)$$

$$\frac{\delta \bar{\rho}(x')}{\delta \rho(r)} = \int_{x_1}^{x_2} \int_{y_1}^{y_2} \int_{z_1}^{z_2} dr'' \frac{\delta \rho(r'')}{\delta \rho(r)} \delta(x' - x'') = \delta(x' - x) \quad (A17)$$

Similarly, the third term in (A12) evaluates to:

$$\frac{\delta}{\delta \rho(r)} \frac{\int_{z_1}^{z'} dz'' \rho(x',y',z'')}{\bar{\rho}(x',y')} = \frac{1}{\bar{\rho}(x',y')} \int_{z_1}^{z'} dz'' \frac{\delta \rho(x',y',z'')}{\delta \rho(r)} - \frac{\int_{z_1}^{z'} dz'' \rho(x',y',z'')}{\bar{\rho}^2(x',y')} \frac{\delta \bar{\rho}(x',y')}{\delta \rho(r)} \quad (A18)$$

The functional derivative in the first term is:

$$\int_{z_1}^{z'} dz'' \frac{\delta \rho(x',y',z'')}{\delta \rho(r)} = \int_{x_1}^{x_2} \int_{y_1}^{y_2} \int_{z_1}^{z'} dr'' \frac{\delta \rho(r'')}{\delta \rho(r)} \delta(x' - x'') \delta(y' - y'') = \delta(x' - x) \delta(y' - y) \Theta(z' - z) \quad (A19)$$

The functional derivative in the second term is:

$$\frac{\delta \bar{\rho}(x',y')}{\delta \rho(r)} = \int_{x_1}^{x_2} \int_{y_1}^{y_2} \int_{z_1}^{z_2} dr'' \frac{\delta \rho(r'')}{\delta \rho(r)} \delta(x' - x'') \delta(y' - y'') = \delta(x' - x) \delta(y' - y) \quad (A20)$$

The functional derivative of the phase factor is then:

$$\frac{\delta \xi_{ji}(r')}{\delta \rho(r)} = \frac{2\pi k_{ji}^x}{N} \Theta(x' - x) + 2\pi k_{ji}^y \left( \frac{\delta(x' - x) \Theta(y' - y)}{\bar{\rho}(x')} - \frac{\int_{y_1}^{y'} dy'' \bar{\rho}(x',y'')}{\bar{\rho}^2(x')} \delta(x' - x) \right)$$

$$+ 2\pi k_{ji}^z \left( \frac{\delta(x' - x) \delta(y' - y) \Theta(z' - z)}{\bar{\rho}(x',y')} - \frac{\int_{z_1}^{z'} dz'' \rho(x',y',z'')}{\bar{\rho}^2(x',y')} \delta(x' - x) \delta(y' - y) \right) \quad (A21)$$

## Appendix B    Density and Exchange-Correlation Matrix Elements

In this section, the density and exchange-correlation matrix elements are derived assuming a constrained basis as in (13). The many-body density can be written as a function of the Kohn-Sham density and the one-body reduced density matrix. Start with the second quantized wavefunction for $N$ particles in $M$ states:

$$|\Psi\rangle = \sum_{\alpha=1}^{P} c_\alpha |\vec{n}^\alpha\rangle \quad (B1)$$

$P \equiv \frac{M!}{N!(M-N)!}$ and $\sum_{\alpha=1}^{P} |c_\alpha|^2 = 1$. Note that sums over Roman letters are sums over $M$ single-particle basis states while sums over Greek letters are sums over $P$ terms in the many-body wavefunction. The density operator in the second quantized formulation is:

$$\hat{\rho} = \sum_{ij=1}^{M} \phi_i^*(r) \phi_j(r) \hat{a}_i^\dagger \hat{a}_j \quad (B2)$$

The many-body density is then

$$\rho(r) = \langle \Psi | \hat{\rho} | \Psi \rangle = \sum_{ij=1}^{M} \phi_i^*(r) \phi_j(r) \rho_{ij} \tag{B3}$$

Rewrite this expression by separating the diagonal and off-diagonal terms in the wavefunction:

$$\rho(r) = \sum_{i=1}^{M} \phi_i^*(r) \phi_i(r) \rho_{ii} + \sum_{i \neq j}^{M} \phi_i^*(r) \phi_j(r) \rho_{ij} \tag{B4}$$

Assuming a constrained basis of the form (13), the diagonal terms can be simplified substituting $\sum_{i=1}^{M} \rho_{ii} = N$ and $|\phi_i(r)|^2 = \frac{1}{N} \rho^{KS}(r)$.

$$\sum_{i=1}^{M} \phi_i^*(r) \phi_i(r) \rho_{ii} = \rho^{KS}(r) \tag{B5}$$

The off-diagonal terms do not vanish identically.

$$\sum_{i \neq j}^{M} \phi_i^*(r) \phi_j(r) \rho_{ij} = \frac{\rho^{KS}(r)}{N} \sum_{i \neq j}^{M} e^{i(\xi_j - \xi_i)} \rho_{ij} \tag{B6}$$

The many-body density can then be written in the following form:

$$\rho(r) = (1 + \Delta\rho(r))\rho^{KS}(r) \tag{B7}$$

The correction due to the off-diagonal terms is defined as:

$$\Delta\rho(r) \equiv \frac{2}{N} \sum_{i<j}^{M} Re\left(\rho_{ij} e^{i(\xi_j(r) - \xi_i(r))}\right) \tag{B8}$$

Note that the one-body reduced density matrix is Hermitian, $\rho_{ij} = \rho_{ji}^*$. The many-body correction to the density (B8) will vanish in the limit as the exchange-correlation potential is exact.

Here the exchange-correlation potential is derived first for the generalized form of the constrained basis (13) and then assuming the ZM basis described in Appendix A as the single particle basis. In order to construct the many-body corrected exchange-correlation potential, set the KS energy equal to the many-body energy (8) and solve for $E_{xc}$:

$$E_{KS} = T_{KS} + E_{ext} + E_H + E_{xc} = \mathcal{E} \tag{B9}$$

$$E_{xc} = \mathcal{E} - T_{KS} - E_{ext} - E_H \tag{B10}$$

$T_{KS}$ is the kinetic energy, $E_{ext}$ is the energy due to external interactions and $E_H$ is the Hartree energy of the non-interacting Kohn Sham (KS) wavefunctions, which can be calculated directly in a DFT framework. By following the usual variational approach to deriving the KS equation where the functional derivative of the KS energy wrt the KS auxiliary wavefunctions, $\phi_i$, is set to zero. The wavefunctions are constrained to be normalized by the method of undetermined multipliers:

$$\frac{\delta}{\delta \phi_i^*}[E_{KS} - \lambda_i(\langle \phi_i^* | \phi_i \rangle - 1)] = 0 \tag{B11}$$

Plugging in the expression for the KS energy (B9) yields:

$$\frac{\delta T_{KS}}{\delta \phi_i^*} + \frac{\delta E_{ext}}{\delta \phi_i^*} + \frac{\delta E_{Hartree}}{\delta \phi_i^*} + \frac{\delta E_{xc}}{\delta \phi_i^*} - \lambda_i \phi_i = 0 \tag{B12}$$

Noting that $\frac{\delta \rho}{\delta \phi_i^*} = \phi_i$, the following components of the potential can be identified including the exchange-correlation potential $v_{xc} \equiv \frac{\delta E_{xc}}{\delta \rho}$ and thus the fourth term above is $\frac{\delta E_{xc}}{\delta \phi_i^*} = \frac{\delta E_{xc}}{\delta \rho} \frac{\delta \rho}{\delta \phi_i^*} = v_{xc} \phi_i$. Similarly, the Hartree potential is defined as $v_H \equiv \frac{\delta E_{Hartree}}{\delta \rho} = \int dr' \frac{\rho(r')}{|r-r'|}$ and the one-body external potential is $v_{ext}^{(1)} = \frac{\delta E_{ext}}{\delta \rho}$. Finally, the Lagrange multipliers are the KS eigenvalues, $\varepsilon_i \equiv \lambda_i$. With these substitutions, (B12) yields the familiar KS equation:

$$-\frac{1}{2}\nabla^2 \phi_i + \left(v_{ext}^{(1)}(r) + v_H(r) + v_{xc}(r)\right)\phi_i = \varepsilon_i \phi_i \tag{B13}$$

Where atomic units were chosen such that $\hbar = m_e = 1$. The Hamiltonian can be defined as

$$\hat{H} = -\frac{1}{2}\nabla^2 + v_{ext}^{(1)}(\boldsymbol{r}) + v_H(\boldsymbol{r}) + v_{xc}(\boldsymbol{r}) \tag{B14}$$

Given the expression for $E_{xc}$ in (B10), the exchange-correlation potential is then:

$$v_{xc}\phi_i = \frac{\delta \mathcal{E}}{\delta \phi_i^*} + \frac{1}{2}\nabla^2 \phi_i - v_{ext}^{(1)}\phi_i - v_H \phi_i \tag{B15}$$

To construct the exchange-correlation potential the functional derivative of the many-body energy must be evaluated

$$\frac{\delta \mathcal{E}}{\delta \phi_i^*} = \frac{\delta T}{\delta \phi_i^*} + \frac{\delta E_{ext}}{\delta \phi_i^*} + \frac{\delta E_{ee}}{\delta \phi_i^*} \tag{B16}$$

Note these are many-body terms from (8), not Kohn-Sham energy terms. The many-body kinetic energy is given by:

$$T = \sum_{k,j=1}^{M} \rho_{kj} t_{kj} \tag{B17}$$

Where the many-body kinetic energy matrix element is defined as:

$$t_{kj} = \left\langle \phi_k \left| -\frac{1}{2}\nabla^2 \right| \phi_j \right\rangle \tag{B18}$$

The functional derivative of the many-body kinetic energy is then:

$$\frac{\delta T}{\delta \phi_i^*} = \sum_{k,j=1}^{M} \rho_{kj} \frac{\delta t_{kj}}{\delta \phi_i^*} \tag{B19}$$

This evaluates trivially to $\frac{\delta T}{\delta \phi_i^*} = -\frac{1}{2}\sum_{j=1}^{M} \rho_{ij} \nabla^2 \phi_j$. This expression results in an orbital-dependent exchange-correlation potential as can be observed when substituting (B16) into (B15).

Next evaluate the second term on the rhs of (B16)

$$\frac{\delta E_{ext}}{\delta \phi_k^*} = \sum_{i,j=1}^{M} \rho_{ij} \frac{\delta v_{ij}^{ext}}{\delta \phi_k^*} \tag{B20}$$

The external potential matrix element for a one-body external potential, $v_{ext}^{(1)}$, is:

$$v_{ij}^{ext} = \left\langle \phi_i \left| v_{ext}^{(1)} \right| \phi_j \right\rangle \tag{B21}$$

The functional derivative in (B20) can be evaluated trivially:

$$\frac{\delta E_{ext}}{\delta \phi_k^*} = \sum_{i,j=1}^{M} \rho_{ij} \delta_{ik} v_{ext}^{(1)}(\boldsymbol{r}) \phi_j(\boldsymbol{r}) = v_{ext}^{(1)}(\boldsymbol{r}) \sum_{j=1}^{M} \rho_{kj} \phi_j(\boldsymbol{r}) \tag{B22}$$

Alternatively, because the one-body matrix element, $v_{ij}^{ext}$, can be written in terms of the density explicitly assuming the density constrained basis (13), the chain rule can be applied to (B20):

$$\frac{\delta E_{ext}}{\delta \phi_k^*} = \sum_{i,j=1}^{M} \rho_{ij} \frac{\delta v_{ij}^{ext}}{\delta \rho} \frac{\delta \rho}{\delta \phi_k^*} = v_{ext} \phi_k \tag{B23}$$

Where the many-body external potential is defined as:

$$v_{ext}(\boldsymbol{r}) \equiv \sum_{i,j=1}^{M} \rho_{ij} \frac{\delta v_{ij}^{ext}}{\delta \rho(\boldsymbol{r})} \tag{B24}$$

To evaluate the potential start with the expression for the $v_{ij}^{ext}$ matrix element assuming the generalized form (13):

$$v_{ij}^{ext} = \frac{1}{N}\int d\boldsymbol{r}' \rho(\boldsymbol{r}') e^{i\xi_{ji}(\boldsymbol{r}')} v_{ext}^{(1)}(\boldsymbol{r}') \tag{B25}$$

Define $\xi_{ji}(\boldsymbol{r}) \equiv \xi_j(\boldsymbol{r}) - \xi_i(\boldsymbol{r})$. To evaluate the functional derivative apply the product rule:

$$\frac{\delta v_{ij}^{ext}}{\delta \rho} = \frac{1}{N}\int d\boldsymbol{r}' \left( \frac{\delta \rho(\boldsymbol{r}')}{\delta \rho(\boldsymbol{r})} e^{i\xi_{ji}(\boldsymbol{r}')} + \rho(\boldsymbol{r}') \frac{\delta e^{i\xi_{ji}(\boldsymbol{r}')}}{\delta \rho(\boldsymbol{r})} \right) v_{ext}^{(1)}(\boldsymbol{r}') \tag{B26}$$

Noting that $\frac{\delta\rho(r')}{\delta\rho(r)} = \delta(r'-r)$

$$\frac{\delta v_{ij}^{ext}}{\delta\rho(r)} = \frac{1}{N}e^{i\xi_{ji}(r)}v_{ext}^{(1)}(r) + \int dr' \rho(r')e^{i\xi_{ji}(r')}\frac{\delta\xi_{ji}(r')}{\delta\rho(r)}v_{ext}^{(1)}(r') \tag{B27}$$

To evaluate the functional derivative in the second term requires an explicit form of $\xi_{ji}(r')$. In the ZM basis, this matrix element can be evaluated substituting in (A21)

$$\frac{\delta v_{ij}^{ext}}{\delta\rho(r)} = \frac{1}{N}e^{i\xi_{ji}^{ZM}(r)}v_{ext}^{(1)}(r) + \frac{2\pi k_{ji}^x}{N}\int_{x_1}^{x}dx'\int_{y_1}^{y_2}dy'\int_{z_1}^{z_2}dz' \rho(x',y',z')e^{i\xi_{ji}^{ZM}(x',y',z')}v_{ext}^{(1)}(x,y',z')$$

$$+ 2\pi k_{ji}^y\left(\frac{1}{\bar{\rho}(x)}\int_{y_1}^{y}dy'\int_{z_1}^{z_2}dz' \rho(x,y',z')e^{i\xi_{ji}^{ZM}(x,y',z')}v_{ext}^{(1)}(x,y',z') - \frac{1}{\bar{\rho}^2(x)}\int_{y_1}^{y_2}dy'\int_{y_1}^{y'}dy''\,\bar{\rho}(x,y'')\int_{z_1}^{z_2}dz' \rho(x,y',z')e^{i\xi_{ji}^{ZM}(x,y',z')}v_{ext}^{(1)}(x,y',z')\right)$$

$$+ 2\pi k_{ji}^z\left(\frac{1}{\bar{\rho}(x,y)}\int_{z_1}^{z}dz' \rho(x,y,z')e^{i\xi_{ji}^{ZM}(x,y,z')}v_{ext}^{(1)}(x,y,z') - \frac{1}{\bar{\rho}^2(x,y)}\int dz' \rho(x,y,z')e^{i\xi_{ji}^{ZM}(x,y,z')}v_{ext}^{(1)}(x,y,z')\int_{z_1}^{z'}dz''\,\rho(x,y,z'')\right) \tag{B28}$$

Substituting back into (B24) and splitting into diagonal ($i=j$) and off-diagonal ($i \neq j$) components over the sum:

$$v_{ext}(r) = v_{ext}^{(1)}(r) + \frac{1}{N}\sum_{i\neq j}^{M}\rho_{ij}\frac{\delta\hat{v}_{ij}^{ext}}{\delta\rho(r)} \tag{B29}$$

The sum over the diagonal components results in a factor of $N$. It can be seen that the diagonal component of the many-body external potential is equal to the KS external potential, which is equal to the one-body external potential. The off-diagonal component is therefore due to correlation effects and is included in the full expression for the exchange-correlation potential. It should be noted that the external potential was assumed to be local in this derivation. For the case of non-local external potentials (e.g. non-local pseudopotential representations of nuclei), the expressions both in (B25) and (B28) must be modified accordingly.

Finally, consider the third term on the rhs of (B16).

$$\frac{\delta E_{ee}}{\delta\phi_m^*} = \frac{1}{2}\sum_{i,j,k,l=1}^{M}\frac{\delta v_{ijkl}^{ee}}{\delta\phi_m^*}\Gamma_{ijkl} \tag{B30}$$

The electron-electron matrix element $v_{ijkl}^{ee}$ is:

$$v_{ijkl}^{ee} = \iint dr' dr'' \frac{\phi_i^*(r')\phi_k^*(r'')\phi_j(r')\phi_l(r'')}{|r'-r''|} \tag{B31}$$

The functional derivative of this matrix element is trivial to evaluate given that $\frac{\delta\phi_i^*(r')}{\delta\phi_j^*(r)} = \delta_{ij}\delta(r-r')$:

$$\frac{\delta v_{ijkl}^{ee}}{\delta\phi_m^*} = \delta_{im}\int dr'' \frac{\phi_k^*(r'')\phi_j(r)\phi_l(r'')}{|r-r''|} + \delta_{jm}\int dr' \frac{\phi_i^*(r')\phi_j(r')\phi_l(r)}{|r'-r|} \tag{B32}$$

Since the sum is over all $i,j$ and the variables of integration are arbitrary, the functional derivative of the electron-electron energy (B30) is

$$\frac{\delta E_{ee}}{\delta\phi_m^*(r)} = \sum_{j,k,l=1}^{M}\Gamma_{mjkl}\left(\int dr' \frac{\phi_k^*(r')\phi_l(r')}{|r-r'|}\right)\phi_j(r) \tag{B33}$$

Alternatively, apply the chain rule so that the functional derivative of the matrix element $v_{ijkl}^{ee}$ is taken wrt the density

$$\frac{\delta E_{ee}}{\delta\phi_m^*} = \frac{1}{2}\sum_{i,j,k,l=1}^{M}\Gamma_{ijkl}\frac{\delta v_{ijkl}^{ee}}{\delta\rho}\frac{\delta\rho}{\delta\phi_m^*} \tag{B34}$$

$$\frac{\delta E_{ee}}{\delta\phi_m^*(r)} = v^{ee}(r)\phi_m(r) \tag{B35}$$

$$v^{ee}(r) = \frac{1}{2}\sum_{i,j,k,l=1}^{M}\Gamma_{ijkl}\frac{\delta v_{ijkl}^{ee}}{\delta\rho} \tag{B36}$$

Where the matrix element can be expressed explicitly in terms of the density assuming the generalized constrained basis from (13)

$$v_{ijkl}^{ee}[\rho] = \frac{1}{N^2} \iint d\mathbf{r}' d\mathbf{r}'' \frac{\rho(\mathbf{r}')e^{i\xi_{ji}(\mathbf{r}')}\rho(\mathbf{r}'')e^{i\xi_{lk}(\mathbf{r}'')}}{|\mathbf{r}'-\mathbf{r}''|} \tag{B37}$$

The functional derivative wrt the density

$$\frac{\delta v_{ijkl}^{ee}}{\delta \rho(\mathbf{r})} = \frac{1}{N^2} \iint d\mathbf{r}' d\mathbf{r}'' \frac{\delta}{\delta \rho(\mathbf{r})}\left(\frac{\rho(\mathbf{r}')e^{i\xi_{ji}(\mathbf{r}')}\rho(\mathbf{r}'')e^{i\xi_{lk}(\mathbf{r}'')}}{|\mathbf{r}'-\mathbf{r}''|}\right) \tag{B38}$$

Applying the product rule results in four terms:

$$\frac{\delta v_{ijkl}^{ee}}{\delta \rho(\mathbf{r})} = \frac{1}{N^2} \iint d\mathbf{r}' d\mathbf{r}'' \left[\frac{\delta \rho(\mathbf{r}')}{\delta \rho(\mathbf{r})}\left(\frac{e^{i\xi_{ji}(\mathbf{r}')}\rho(\mathbf{r}'')e^{i\xi_{lk}(\mathbf{r}'')}}{|\mathbf{r}'-\mathbf{r}''|}\right) + \left(\frac{\rho(\mathbf{r}')e^{i\xi_{ji}(\mathbf{r}')}\rho(\mathbf{r}'')e^{i\xi_{lk}(\mathbf{r}'')}}{|\mathbf{r}'-\mathbf{r}''|}\right)\frac{\delta \xi_{ji}(\mathbf{r}')}{\delta \rho(\mathbf{r})} \right.$$
$$\left. + \frac{\delta \rho(\mathbf{r}'')}{\delta \rho(\mathbf{r})}\left(\frac{\rho(\mathbf{r}')e^{i\xi_{ji}(\mathbf{r}')}e^{i\xi_{lk}(\mathbf{r}'')}}{|\mathbf{r}'-\mathbf{r}''|}\right) + \left(\frac{\rho(\mathbf{r}')e^{i\xi_{ji}(\mathbf{r}')}\rho(\mathbf{r}'')e^{i\xi_{lk}(\mathbf{r}'')}}{|\mathbf{r}'-\mathbf{r}''|}\right)\frac{\delta \xi_{lk}(\mathbf{r}'')}{\delta \rho(\mathbf{r})}\right] \tag{B39}$$

The first and third terms are trivial to evaluate given that $\frac{\delta \rho(\mathbf{r}')}{\delta \rho(\mathbf{r})} = \delta(\mathbf{r}' - \mathbf{r})$ and $\frac{\delta \rho(\mathbf{r}'')}{\delta \rho(\mathbf{r})} = \delta(\mathbf{r}'' - \mathbf{r})$.

$$\iint d\mathbf{r}' d\mathbf{r}'' \frac{\delta \rho(\mathbf{r}')}{\delta \rho(\mathbf{r})}\left(\frac{e^{i\xi_{ji}(\mathbf{r}')}\rho(\mathbf{r}'')e^{i\xi_{lk}(\mathbf{r}'')}}{|\mathbf{r}'-\mathbf{r}''|}\right) = \int d\mathbf{r}'' \frac{e^{i\xi_{ji}(\mathbf{r})}\rho(\mathbf{r}'')e^{i\xi_{lk}(\mathbf{r}'')}}{|\mathbf{r}-\mathbf{r}''|} \tag{B40}$$

$$\iint d\mathbf{r}' d\mathbf{r}'' \frac{\delta \rho(\mathbf{r}'')}{\delta \rho(\mathbf{r})}\left(\frac{\rho(\mathbf{r}')e^{i\xi_{ji}(\mathbf{r}')}e^{i\xi_{lk}(\mathbf{r}'')}}{|\mathbf{r}'-\mathbf{r}''|}\right) = \int d\mathbf{r}' \frac{\rho(\mathbf{r}')e^{i\xi_{ji}(\mathbf{r}')}e^{i\xi_{lk}(\mathbf{r})}}{|\mathbf{r}'-\mathbf{r}|} \tag{B41}$$

Because the variables of integration are arbitrary, the first term with indices $ijkl$ is equal to the third term with indices $klij$ and thus the sum of the first and third terms are equal. To evaluate the second and fourth terms requires an explicit form of the phase factor. For the ZM basis substitute (A21) into the second term:

$$\iint d\mathbf{r}' d\mathbf{r}'' \left(\frac{\rho(\mathbf{r}')e^{i\xi_{ji}(\mathbf{r}')}\rho(\mathbf{r}'')e^{i\xi_{lk}(\mathbf{r}'')}}{|\mathbf{r}'-\mathbf{r}''|}\right)\frac{\delta \xi_{ji}(\mathbf{r}')}{\delta \rho(\mathbf{r})}$$
$$= \frac{2\pi k_{ji}^x}{N}\int_{x_1}^x dx' \int_{y_1}^{y_2} dy' \int_{z_1}^{z_2} dz' \int d\mathbf{r}'' \left(\frac{\rho(\mathbf{r}')e^{i\xi_{ji}(\mathbf{r}')}\rho(\mathbf{r}'')e^{i\xi_{lk}(\mathbf{r}'')}}{|\mathbf{r}'-\mathbf{r}''|}\right) + \frac{2\pi k_{ji}^y}{\bar{\rho}(x)}\int_{y_1}^y dy' \int_{z_1}^{z_2} dz' \int d\mathbf{r}'' \left(\frac{\rho(x,y',z')e^{i\xi_{ji}(x,y',z')}\rho(\mathbf{r}'')e^{i\xi_{lk}(\mathbf{r}'')}}{|(x,y',z')-\mathbf{r}''|}\right)$$
$$- \frac{2\pi k_{ji}^y}{\bar{\rho}^2(x)}\int_{y_1}^{y_2} dy' \int_{z_1}^{z_2} dz' \int d\mathbf{r}'' \frac{\rho(x,y',z')e^{i\xi_{ji}(x,y',z')}\rho(\mathbf{r}'')e^{i\xi_{lk}(\mathbf{r}'')}}{|(x,y',z')-\mathbf{r}''|}\int_{y_1}^{y'} dy'' \bar{\rho}(x,y'') + \frac{2\pi k_{ji}^z}{\bar{\rho}(x,y)}\int_{z_1}^z dz' \int d\mathbf{r}'' \frac{\rho(x,y,z')e^{i\xi_{ji}(x,y,z')}\rho(\mathbf{r}'')e^{i\xi_{lk}(\mathbf{r}'')}}{|(x,y,z')-\mathbf{r}''|}$$
$$- \frac{2\pi k_{ji}^z}{\bar{\rho}^2(x,y)}\int_{z_1}^{z_2} dz' \int d\mathbf{r}'' \frac{\rho(x,y,z')e^{i\xi_{ji}(x,y,z')}\rho(\mathbf{r}'')e^{i\xi_{lk}(\mathbf{r}'')}}{|(x,y,z')-\mathbf{r}''|}\int_{z_1}^{z'} dz'' \rho(x,y,z'') \tag{B42}$$

The fourth term with indices $ijkl$ is equal to the second term with indices $klij$. Given the symmetry over the indices, the functional derivative of the electron-electron energy for the ZM basis can be written as:

$$\frac{\delta v_{ijkl}^{ee}}{\delta \rho(\mathbf{r})} = \mathbb{V}_{ijkl}^{ee}(\mathbf{r}) + \mathbb{V}_{klij}^{ee}(\mathbf{r}) \tag{B43}$$

The function $\mathbb{V}_{ijkl}^{ee}(\mathbf{r})$ is defined:

$$\mathbb{V}^{ee}_{ijkl}(\boldsymbol{r}) \equiv e^{i\xi_{ji}(\boldsymbol{r})}\int d\boldsymbol{r}'' \frac{\rho(\boldsymbol{r}'')e^{i\xi_{lk}(\boldsymbol{r}'')}}{|\boldsymbol{r}-\boldsymbol{r}''|} + \frac{2\pi k^x_{ji}}{N}\int_{x_1}^{x} dx' \int_{y_1}^{y_2} dy' \int_{z_1}^{z_2} dz' \int d\boldsymbol{r}'' \left( \frac{\rho(\boldsymbol{r}')e^{i\xi_{ji}(\boldsymbol{r}')}\rho(\boldsymbol{r}'')e^{i\xi_{lk}(\boldsymbol{r}'')}}{|\boldsymbol{r}'-\boldsymbol{r}''|} \right)$$
$$+ \frac{2\pi k^y_{ji}}{\bar{\rho}(x)} \int_{y_1}^{y} dy' \int_{z_1}^{z_2} dz' \int d\boldsymbol{r}'' \left( \frac{\rho(x,y',z')e^{i\xi_{ji}(x,y',z')}\rho(\boldsymbol{r}'')e^{i\xi_{lk}(\boldsymbol{r}'')}}{|(x,y',z')-\boldsymbol{r}''|} \right)$$
$$- \frac{2\pi k^y_{ji}}{\bar{\rho}^2(x)} \int_{y_1}^{y_2} dy' \int_{z_1}^{z_2} dz' \int d\boldsymbol{r}'' \frac{\rho(x,y',z')e^{i\xi_{ji}(x,y',z')}\rho(\boldsymbol{r}'')e^{i\xi_{lk}(\boldsymbol{r}'')}}{|(x,y',z')-\boldsymbol{r}''|} \int_{y_1}^{y'} dy'' \bar{\rho}(x,y'')$$
$$+ \frac{2\pi k^z_{ji}}{\bar{\rho}(x,y)} \int_{z_1}^{z} dz' \int d\boldsymbol{r}'' \frac{\rho(x,y,z')e^{i\xi_{ji}(x,y,z')}\rho(\boldsymbol{r}'')e^{i\xi_{lk}(\boldsymbol{r}'')}}{|(x,y,z')-\boldsymbol{r}''|}$$
$$- \frac{2\pi k^z_{ji}}{\bar{\rho}^2(x,y)} \int_{z_1}^{z_2} dz' \int d\boldsymbol{r}'' \frac{\rho(x,y,z')e^{i\xi_{ji}(x,y,z')}\rho(\boldsymbol{r}'')e^{i\xi_{lk}(\boldsymbol{r}'')}}{|(x,y,z')-\boldsymbol{r}''|} \int_{z_1}^{z'} dz'' \rho(x,y,z'')$$
(B44)

For the case where $i = j$ and $k = l$, the functional derivative above reduces to the Hartree potential:

$$\frac{\delta v^{ee}_{iikk}}{\delta \rho(\boldsymbol{r})} = \frac{2}{N^2} \int d\boldsymbol{r}' \frac{\rho(\boldsymbol{r}')}{|\boldsymbol{r}'-\boldsymbol{r}|} = \frac{2}{N^2} v_H(\boldsymbol{r}) \tag{B45}$$

The many-body electron-electron potential (B36) can be written as the sum of terms diagonal in the two pairs $i = j$ and $k = l$ and the off-diagonal components:

$$v_{ee}(\boldsymbol{r}) = \frac{v_H(\boldsymbol{r})}{N^2} \sum_{i,k=1}^{M} \Gamma_{iikk} + \frac{1}{2} \sum_{i\neq j, k\neq l=1}^{M} \frac{\delta v^{ee}_{ijkl}}{\delta \rho} \Gamma_{ijkl} \tag{B46}$$

For $N$ electrons in $M$ states, it can be shown that $\sum_{i,k=1}^{M} \Gamma_{iikk} = N(N-1)$. Substituting this into the above equation leaves the following expression for the many-body electron-electron potential:

$$v_{ee}(\boldsymbol{r}) = v_H(\boldsymbol{r})\left(1 - \frac{1}{N}\right) + \frac{1}{2}\sum_{i\neq j, k\neq l}^{M} \frac{\delta v^{ee}_{ijkl}}{\delta \rho} \Gamma_{ijkl} \tag{B47}$$

The many-body corrected exchange-correlation operator applied to state $\phi_m$ can then be obtained by substituting (B16), (B19), (B29), and (B47) into (B15)

$$v_{xc}\phi_m = \frac{1}{2}\sum_{j=1}^{M}(\delta_{mj} - \rho_{mj})\nabla^2 \phi_j + v^{loc}_{xc}(\boldsymbol{r})\phi_m \tag{B48}$$

$v^{loc}_{xc}$ is the local exchange-correlation potential (again, assuming a local external potential) for the ZM basis:

$$v^{loc}_{xc}(\boldsymbol{r}) \equiv \frac{1}{N}\sum_{i\neq j}^{M} \rho_{ij} \frac{\delta v^{ext}_{ij}}{\delta \rho} + \frac{1}{2}\sum_{i\neq j, k\neq l=1}^{M} \frac{\delta v^{ee}_{ijkl}}{\delta \rho} \Gamma_{ijkl} - \frac{1}{N} v_H(\boldsymbol{r}) \tag{B49}$$

The matrix elements $\frac{\delta v^{ext}_{ij}}{\delta \rho}$ and $\frac{\delta v^{ee}_{ijkl}}{\delta \rho}$ are basis/density dependent and can be calculated according to (B28) and (B43). The many-body corrected Hamiltonian operator applied to a state $\phi_m$ is:

$$\hat{H}|\phi_m\rangle = -\frac{1}{2}\sum_{j=1}^{M} \rho_{mj}\nabla^2|\phi_j\rangle + \hat{v}^{(1)}_{ext}|\phi_m\rangle + \hat{v}_H|\phi_m\rangle + \hat{v}^{loc}_{xc}|\phi_m\rangle \tag{B50}$$

Note that that only the first term depends on the basis. The remaining terms are local and basis independent and can therefore be applied to an arbitrary state in the DFT framework. In order to apply the first term to an arbitrary state, $|\psi\rangle$, in the DFT framework it is necessary to expand in the basis of the constructed basis $|\psi\rangle = \sum_m |\phi_m\rangle\langle\phi_m|\psi\rangle$:

$$\hat{H}|\psi\rangle = -\frac{1}{2}\sum_{m,j=1}^{M} \rho_{mj}\langle\phi_m|\psi\rangle\nabla^2|\phi_j\rangle + \hat{v}^{(1)}_{ext}|\psi\rangle + \hat{v}_H|\psi\rangle + \hat{v}^{loc}_{xc}|\psi\rangle \tag{B51}$$